# Transformer-Based Active Learning for Data-Efficient Vaccine Epitope Selection in PRRS

Aspen Erlandsson Brisebois[1,2], Zahed Khatooni[1,3,4], Connor Burbridge[5], Brook Byrns[5], Heather L. Wilson[1,6], Sureesh Tikoo[1,7*], Steven Rayan[3,4*], Gordon Broderick[1,3,4*]

[1] Vaccine and Infectious Disease Organization (VIDO), University of Saskatchewan, Saskatoon, SK, Canada

[2] Engineering Science, Faculty of Applied Science, University of Toronto, Toronto, ON, Canada

[3] Department of Mathematics and Statistics, University of Saskatchewan, Saskatoon, SK, Canada

[4] Centre for Quantum Topology and Its Applications (quanTA), University of Saskatchewan, Saskatoon, SK, Canada

[5] Information and Communications Technology, University of Saskatchewan, Saskatoon SK, Canada

[6] Department of Veterinary Microbiology, Western College of Veterinary Medicine, University of Saskatchewan, Saskatoon, Saskatchewan, Canada, S7N 5B4

[7] School of Public Health, Vaccinology & Immunotherapeutics program, University of Saskatchewan, Saskatoon, Saskatchewan, Canada, S7N 5E3

* Corresponding authors: : sureesh.tikoo@usask.ca; steven.rayan@usask.ca ; gordon.broderick@usask.ca

## Abstract

High-fidelity molecular docking simulations can produce biologically relevant estimates of epitope-receptor binding affinity but are computationally expensive and therefore limit the number of candidates that can be screened for vaccine design. In this work, we evaluate machine learning (ML) approaches where variants of active learning are used to classify instances of high binding affinity between 9-mer epitopes and a well-conserved swine leukocyte antigen (SLA) receptor in the context of Porcine Reproductive and Respiratory Syndrome (PRRS). We use an internally generated dataset of 80 epitope-SLA docking affinities, each requiring >48 hours of high-performance computing (HPC). Multiple model families (linear, MLP, CNN, and a small transformer) are trained under strict low-data conditions within a pool-based active learning loop. In each case optimal model configurations are identified by conducting large-scale hyperparameter optimization over the combined space of model architecture, training configuration, acquisition policy, and ensemble decision rules. To mitigate the effects of data subsample selection, each candidate configuration is evaluated by averaging performance over many randomized and balanced training and validation data subsets. Across experiments, transformer-based sequence models consistently emerged as the best performing architecture, with active incremental learning yielding significant improvement over a baseline random sample acquisition strategy. Under moderate training data availability (N=30), the optimized ML-model configuration outperforms a standard baseline trained on twice the amount of data. Under higher training data availability(N=60) the same configuration achieves a peak accuracy of 86.8%, consistent with an upper bound of 85% classification accuracy based on two independent estimates of conformational noise.

## 1. Introduction

Computational vaccine design faces an unfavorable trade-off between biological fidelity and throughput. While molecular docking and related simulation approaches can produce estimates of protein-protein binding affinity that are directly relevant to prioritizing epitope candidates, they are dauntingly slow and resource intensive to apply at scale. This bottleneck is especially limiting when the candidate space is large (often thousands to tens of thousands of epitopes per pathogen). In this work we propose a broadly applicable approach for the rapid screening of candidate vaccine antigens based on the accurate classification of binding affinity for candidate epitopes from the porcine reproductive and respiratory syndrome (PRRS) proteome and swine leukocyte antigen (SLA) receptors. A key practical constraint in this setting is the low number of available labeled examples, with each new data point requiring approximately 48 hours of high-performance computing to estimate. Under such conditions, methods that are robust in low-data regimes are not only desirable; they are necessary. Such constraints on data availability become even more critical in fast-response scenarios such as emerging outbreaks or pandemic conditions, when decision timelines are compressed and the candidate epitope space can be large. In such cases, the ability to rapidly prioritize epitopes for follow-up study can be critically important, and computational methods that reduce dependence on exhaustive high-cost simulations are paramount in guiding focused experimental work to reduce overall vaccine development timelines. While PRRS is used here as a case study, the formalism and numerical protocols proposed here have been designed with these broader rapid-response considerations in mind.

Though touted, at least in the media, as requiring almost unlimited amounts of data, recent work has shown that selective incremental machine learning in artificial intelligence (AI) model architectures can deliver robust high-quality predictions using only a small fraction of available data [Guo et al., 2025]. Active learning, a static variant of reinforcement learning, is ideally suited to environments where existing data are limited and where each additional labelled example is expensive to incorporate. Rather than training once using all examples in a fixed dataset, pool-based active learning iteratively selects the next most informative samples from an unlabeled pool according to an acquisition policy. In the numerical experiments presented here, incremental active learning consistently outperforms conventional one-time training on the full data set under identical split of training and validation subsets. This methodology also offers significant operational advantages as small seed data sets can be used initially followed by new optimally informative docking experiments that are generated on demand using the same acquisition policies to expand the dataset with optimal resources usage.

## 2. Methods

The overall proposed framework is composed of multiple components, each with features that can be tailored to a specific problem or environment (**Figure 1**).

### *2.1. Dataset Preparation and Preprocessing*

We used an internally generated dataset of 80 epitope-SLA docking results with each row containing (i) a 9-mer epitope amino-acid sequence, (ii) a scalar docking-derived binding affinity estimate, and (iii) a precomputed binary binding class assignment (Strong/Weak). We denote this original full dataset as $\mathcal{D}_{\text{full}}$. The Strong/Weak class boundary was obtained by applying Otsu thresholding [Otsu, 1979] to the binding affinity distribution, yielding a class cutoff of $\tau_A = -77.8$. We define Strong binders as those with $A < \tau_A$ and Weak binders otherwise. Applying this threshold, the dataset contains 38 Strong and 42 Weak samples. To ensure strict class balance during all experiments, each run uses all 38 Strong samples and a randomly selected subset of 38 of the 42 Weak samples, effectively down sampling the majority class. Therefore, the effective balanced dataset size is 76 samples in each run.

Before each training run is performed a new random split of the dataset is prepared (**Figure 2**). The final important step of dataset preparation is normalizing the binding affinity scalar, $A \in \mathbb{R},\ -108.5 \leq A \leq -48.5$, to a suitable range for ML training. A standard range compatible with the Sigmoid activation function [Dubey et al., 2022] is $\tilde{A} \in \mathbb{R},\ 0 \leq \tilde{A} \leq 1$. We define the normalizer $\mathcal{N}$ as a piecewise linear transform such that $\tilde{A} = \mathcal{N}_\psi A$ where the subscript $\psi$ denotes the range-specific parameters (min, max, threshold). $\mathcal{N}_\psi$ is always fitted to the initial training dataset, $\mathcal{D}_{\text{init}}$ and then applied to the pool and test sets $\mathcal{D}_{\text{pool}}$ , $\mathcal{D}_{\text{test}}$. This step strictly prevents information leakage from the training the pool and test sets. We could have fitted $\mathcal{N}_\psi$ to the entire training set including pool, but maintaining isolation keeps portability to using these methods in situations where the pool is truly unlabeled, for example when expanding the dataset by using the active-learning pipeline for design of experiments.

$$\mathcal{N}_\psi(x) = \text{clip}\left(\begin{cases} 0.5 + 0.5\dfrac{\tau_A - x}{\tau_A - \min(\mathcal{D}_{\text{init}})}, & x < \tau_A \\ 0.5 - 0.5\dfrac{x - \tau_A}{\max(\mathcal{D}_{\text{init}}) - \tau_A}, & x \geq \tau_A \end{cases}, 0, 1\right)$$

Concisely, $\mathcal{N}_\psi(x)$ fits Weak affinities greater than the threshold to between 0 to 0.5, and Strong affinities less than the threshold between 0.5 and 1.0. If the normalization is applied to data outside of the initial min/max range, a linear mapping could potentially place it outside normalization bounds hence we map to a bounded window of [0,1]. To ensure consistent behavior across data subsets we apply this piecewise mapping to the full dataset $\mathcal{D}_{\text{full}}$ (**Figure 3(A)**) and demonstrate that dynamic range is also well maintained in a small random subset of 5+5 samples (**Figure 3(B)**). Each run uses this formal data preparation process to ensure class balancing, random shuffling, and no informational leakage of the test set during normalization. Details of the overall data preparation procedure are illustrated using the example of a N=30 subset, with 11+11 in the initial training set and 4+4 in the pool (**Figure 2**). Pseudo code is presented in **Supplementary Table 1**.

*2.2. Problem Formulation and Learning Targets*

All models in this work are evaluated based on predicted accuracy in a binary classification task consisting of assigning an epitope from a held-out test set to either a Strong or Weak affinity class. While training is conducted using the continuous affinity estimates provided by the docking experiments, the discrete class assignment accuracy is the evaluation metric used to compare model selection and hyperparameter settings. We consider two alternative approaches to supervised training:

a. **Label class training** where the model is trained directly to predict class assignment $y \in \{\text{Weak}, \text{Strong}\} = \{0, 1\}$ using a classification loss i.e. binary cross entropy [Dubey et al., 2022].

b. **Affinity proxy training** where the model is trained to predict a smooth, normalized target $\tilde{A}$, derived from the binding affinity as described in Section 2.1 using a regression loss (Mean Squared Error, Mean Absolute Error, or Huber). [Jadon et al., 2024]. Affinities are mapped onto the interval [0,1] such that the class threshold corresponds exactly to 0.5. During the evaluation of classification accuracy the model predictions, $F(\mathcal{D}_{\text{test}})$, are converted to a predicted class assignment by applying a threshold of 0.5.

We define the training target choice as $\mathcal{T} \in \{\text{class}, \text{affinity}\}$. This choice allows us to assess if a continuous proxy target stabilizes learning or improves acquisition behavior in low data settings. Affinity prediction is only used as a proxy supervision signal and is not evaluated as the objective, only class accuracy is reported.

### *2.3. Input Encoding*

For each sample epitope, we apply a one-hot encoding over the 20 standard amino acid vocabulary of {A,C,D,E,F,G,H,I,K,L,M,N,P,Q,R,S,T,V,W,Y} to convert the sequence of 9 amino acids into a binary array of shape (9,20) (**Figure 4**). The Convolutional Neural Network (CNN) [LeCun et al., 2002] and Multilayer Perceptron (MLP) [Rumelhart et al., 1986] models operate on this representation directly. All four model families operate on this (9,20) array as input. The Linear and Multilayer Perceptron (MLP) [Rumelhart et al., 1986] models flatten this array to a (1,180) vector, while Convolutional Neural Network (CNN) [LeCun et al., 2002] convolves along the 9-position axis. In the case of the transformer mode, each amino acid column vector is treated as a token. From these tokens, a linear projection is learned such that each amino acid 20-dimesional one-hot encoded representation is positioned in an embedding of user-specified dimensionality (d_model).

### *2.4. Model Families and Architecture Search*

The model family and specific structure are critically important to the performance of a given predictor. We open the search space to 4 possible model families, each with their own tunable structural parameters such as layer depth, activation function, dropout, etc...(**Table 1**) The model family choice is defined as:

$$f_\theta \in \{\text{Linear (Lin)}, \text{MLP}, \text{CNN}, \text{Transformer (Tr)}\}$$

With the following providing a summary description of each model family:

- **Linear:** A single linear mapping from the flattened one-hot encoded sequence to a scalar output, with no hidden layers or nonlinear transformations.
- **MLP:** A feedforward neural network composed of one or more fully connected hidden layers with nonlinear activations, operating on the flattened one-hot sequence representation [Rumelhart et al., 1986].
- **CNN:** A one-dimensional convolutional neural network applied along the sequence dimension, consisting of stacked convolution and pooling layers followed by fully connected layers [LeCun et al., 2002].
- **Transformer:** A self-attention-based sequence model that treats epitope positions as tokens, projecting one-hot inputs into an embedding space and applying stacked multi-head attention and feedforward layers [Vaswani et al., 2017].

*2.5. Training Configuration and Optimization*

We optimize over standard training hyperparameters (**Table 2**) including the choice of search algorithm for model weight identification $\sigma$ (e.g. Adam, etc…), the learning rate $L_r$ (log-scaled), learning momentum *M* (in SGD and RMSProp), weight decay $\lambda_{\text{wd}}$, batch size *B*, and the choice of loss function $\mathcal{L}$ conditional on the training target $\mathcal{T}$.

*2.6. Active Learning Framework*

Each numerical experiment presented here requires conducting a pool-based active learning procedure where model training and data acquisition are iteratively interleaved. An initial labeled dataset $\mathcal{D}_{\text{init}}$ is constructed according to the protocol described in Section II.A and the remaining labeled samples withheld as used as an unlabeled pool $\mathcal{D}_{\text{pool}}$. Additionally, a hold-out test set or external validation set $\mathcal{D}_{\text{test}}$ is selected and held constant for the duration of the active learning run and is never used for training, acquisition, or normalization. Models are trained in two phases:

- For the initial epochs, $E_{\text{init}}$ on the initial labeled set $D_{\text{init}}$
- For the step epochs, $E_{\text{step}}$ after each acquisition step as labeled data grows $\mathcal{D}_{\text{train}} = \mathcal{D}_{\text{init}} \cup \mathcal{D}_{\text{added}}$, where $\mathcal{D}_{\text{added}}$ comes from applying the next-sample selection policy applied to $\mathcal{D}_{\text{pool}}$.

At each active learning step, an acquisition policy is applied to score all samples in $\mathcal{D}_{\text{pool}}$, and the top $\mathrm{k}_{\text{step}}$ samples are selected and added to the labeled training set, $\mathcal{D}_{\text{train}}$. The model is then retrained for $\mathrm{E}_{\text{step}}$ epochs on the expanded training set, after which model performance is evaluated against $\mathcal{D}_{\text{test}}$ and the selection of new training samples from $\mathcal{D}_{\text{pool}}$ is repeated. This process continues until the set $\mathcal{D}_{\text{pool}}$, is exhausted (**Figure 5, Supplementary Table 1**). For each numerical experiment the resulting evolution in classification performance on the validation set $\mathcal{D}_{\text{test}}$ is recorded as a function of the increasing number of training samples and these trajectories aggregated across repeated experiments using the same model hyperparameter settings on randomized sample subsets as described in Section II.I.

*2.7. Acquisition Policies*

At each active learning step, an acquisition policy is used to select a subset of samples from the unlabeled pool $D_{\text{pool}}$ for labeling and inclusion in the training set. Let $k_{\text{step}}$ denote the configured acquisition batch size. If $\text{size}(\mathcal{D}_{\text{pool}}) < k_{\text{step}}$, all remaining pool samples are selected, and the active learning loop terminates after that step. We define the acquisition policy choice as

$$\alpha \in \mathcal{A},$$

where the acquisition policy space $\mathcal{A}$ is given by:

$$\mathcal{A} = \{\text{Random}, \text{Least Confident}, \text{K-Center}, \text{EGL}\}$$

For a given selection policy, a corresponding scalar score is computed and every sample $x \in \mathcal{D}_{\text{pool}}$ ranked accordingly with only the top $k_{\text{step}}$ samples selected for inclusion in the augmented labeled training set. In specific sample acquisition policies this score can be based on a more granular intermediate representation of the output namely the feature vector *h(x)* or the model output representation at a layer immediately preceding the final output layer. This more granular intermediate representation is used both as an embedding in coverage-based selection (K-Center) [Sener and Savarese, 2017] and as the feature component in gradient-based acquisition (EGL) [Shukla, 2022].

For probabilistic acquisition policies such as random acquisition or EGL, let $p(x) \in [0,1]$ denote the predicted probability of the Strong affinity class for sample $x$, obtained from a sigmoidal cumulative probability density function. When model ensembles are used, $p(x)$ is defined as the mean predicted probability across ensemble members. Each policy assigns a scalar score $s(x)$ to every sample $x \in \mathcal{D}_{\text{pool}}$, and the top $k_{\text{step}}$ samples according to this score are selected as follows:

- **Random acquisition:** Samples are selected uniformly at random from $\mathcal{D}_{\text{pool}}$, independent of model predictions.
- **Least confident candidates:** Samples are ranked by proximity to the decision boundary,

  $$s(x) = 1 - | p(x) - 0.5 |,$$

  and those with the highest scores (i.e., closest to 0.5) are selected.
- **K-center:** Samples are selected to maximize coverage in a learned feature space. Let the feature vector $h(x)$ denote the granular intermediate representation of the output. For each candidate sample $x \in \mathcal{D}_{\text{pool}}$, we compute its distance *d(x)* in this feature space to the closest sample in the current labeled set $\mathcal{D}_{\text{train}}$,

  $$d(x) = \min_{x' \in \mathcal{D}_{\text{train}}} \| h(x) - h(x') \|.$$

  The sample with the largest value of *s(x)* = $d(x)$is selected and added to the labeled set. This procedure is repeated until $k_{\text{step}}$ samples have been selected.
- **EGL:** A proxy score $s(x)$ for the expected gradient length is computed using the predicted probability and the norm of the granular intermediate feature vector $h(x)$,

  $$s(x) \propto p(x)\,(1 - p(x))\ \| h(x) \|,$$

  Higher scores $s(x)$ favor samples expected to drive larger parameter updates in the final layer.

For ensemble-based acquisition, scores $s_m(x)$ are computed independently for each ensemble member and then averaged prior to ranking.

### *2.8. Ensemble Modeling and Decision Rules*

In addition to single-model training, we evaluate ensemble-based configurations. When ensembles are enabled, multiple models with identical architectures and hyperparameters are instantiated using different random weight initializations and trained in parallel on the same labeled data. Ensembles are used both for evaluation and acquisition scoring where applicable.

Let $\{p_m(x)\}_{m=1}^{M}$denote the predicted probabilities from an ensemble of size $M$. We define the choice of ensemble decision rule as

$$\delta \in D,$$

where

$$D = \{\text{Majority Vote}, \text{Mean Round}, \text{Confidence Weighted}\}$$

A precise definition of each rule follows below where $\mathbb{I}[\cdot]$ denotes the indicator function, which evaluates to 1 if its argument is true and 0 otherwise. The ensemble decision rules included in the optimization search space are summarized in **Table 4**.

- **Majority vote:** Each ensemble member produces a binary prediction $\hat{y}_m(x) = \mathbb{I}[p_m(x) \geq 0.5]$. The final prediction is given by

$$\hat{y}(x) = \mathbb{I}\left[\sum_{m=1}^{M} \hat{y}_m(x) \geq \frac{M}{2}\right].$$

- **Mean round:** The ensemble mean predicted probability

$$\bar{p}(x) = \frac{1}{M}\sum_{m=1}^{M} p_m(x)$$

  A threshold of 0.5 is then applied to $\bar{p}(x)$ in order to produce the final predicted class assignment $\hat{y}(x)$.

$$\hat{y}(x) = \mathbb{I}[\bar{p}(x) \geq 0.5]$$

- **Confidence weighted**: Each binary prediction $\hat{y}_m(x)$ is weighted by the model confidence $w_m(x) = | p_m(x) - 0.5 |$. The final prediction is obtained by thresholding the normalized weighted vote

$$\hat{y}(x) = \mathbb{I}\left[\frac{\sum_{m=1}^{M} w_m(x)\, \hat{y}_m(x)}{\sum_{m=1}^{M} w_m(x)} \geq 0.5\right]$$

### *2.9. Repeated Randomized Evaluation*

Because this framework must be deployable to situations where data may be extremely sparse, we require that hyperparameter estimation be highly robust to the potentially high variability arising from the random selection of train/test subsets as well as the random initialization of model parameters. The goal of this work is to identify policies and architectures that are consistently strong across these unavoidable sources of variability. Accordingly, we treat each configuration as delivering a distribution of outcomes rather than a single point estimate with the performance of every hyperparameter solution set computed over repeated runs. In each run we generate a new class-balanced randomly selected data subsets (including initial/pool composition) and reinitialize model parameters with a distinct random seed. In an attempt to balance thoroughness against runtime, we conducted hyperparameter optimization across 1000 iterations, where each iteration consisted of 40 repeated runs and where each run used its own random initialization seed. In addition to model design, this work seeks to explore and evaluate different methods of next-sample selection. To facilitate this comparison, each training run for a particular model configuration conducts duplicate parallel executions once active learning starts, one with the selection policy of interest, and a second that always uses random selection for comparison.

### *2.10. Hyperparameter Optimization Procedure*

To identify effective combinations of model architecture, training configuration, active learning strategy, and ensemble settings, we perform automated hyperparameter optimization over the full method space defined in the preceding sections. Each candidate configuration specifies a complete pipeline, including the model family and structure, training parameters, active learning parameters, and (when enabled) ensemble settings. The optimization objective is defined as the peak classification accuracy over the test set $\mathcal{D}_{\text{test}}$ achieved over the active learning trajectory. For a given configuration, repeated randomized evaluation (Section 2.9) produces a mean accuracy curve as a function of the number of labeled training samples. The scalar objective value assigned to the configuration is the maximum value attained along this mean trajectory. Hyperparameter optimization is performed using Bayesian optimization routine with a Tree-structured Parzen Estimator (TPE) sampler available in the Python library Optuna [Watanabe, 2023]. The search is conducted for a total of 200 optimization trials, and all trial results are saved to a database backend to allow interruption and resumption of the study without loss of information. To ensure optimization stability and faster convergence, the search space was decoupled by model family: independent optimization studies were executed in parallel for each architecture (Linear, MLP, CNN, Transformer) and the results were aggregated to identify the global best configuration.

### *2.11. Noise Ceiling and Maximum Achievable Accuracy Estimation*

To assess when further methodological complexity is unlikely to yield meaningful improvement in the context of this use case, we estimate a practical upper bound on achievable classification accuracy imposed by the inherent variability in docking scores that arise for the same epitope-receptor pair as a result of competing three-dimensional conformations of a given epitope amino acid sequence. In an attempt to characterize the magnitude of this conformational variability, we analyze internally generated datasets consisting of replicate in silico docking results in a random subset of sequences. Each conformation is generated from this primary sequence using AlphaFold2 [Yang et al., 2023] and binding affinity estimated under identical molecular docking and scoring settings. Using these estimated parameters, we perform a Monte Carlo simulation to estimate the maximum attainable classification accuracy when conformational noise is included in binding affinity. The procedure is as follows:

1. Assume a hypothetical noiseless oracle classifier that always predicts the correct class label based on the true, noiseless affinity.
2. For each epitope, apply conformational noise to the true affinity by sampling from a normal distribution with the estimated standard deviation.
3. Re-apply the fixed classification threshold ($\tau_A$= -77.8) to the noise-adjusted affinity.
4. Count instances where the noise-adjusted label differs from the original label. These represent unavoidable errors even for a perfect classifier.
5. Repeat the process over 1000 trials to estimate the expected accuracy ceiling.

We first apply this protocol to estimate the replicate error in class assignment that arises from the variance in binding affinity in 10 distinct conformations of a single epitope, RYKHTWGFE, bound to the same well-conserved SLA used as the reference receptor throughout this work (**Table 6**). In an attempt to verify if these estimates are representative of the broader set of epitope amino acid sequences we also asses conformational variability in a separate set of 5 randomly selected example sequences each with duplicate estimates of binding affinity as predicted by in silico docking (**Table 7**). The Shapiro-Wilk test is applied to assess normality with Bartlett's test and Levene's test being used to assess homoscedasticity.

*2.12. Computing environment*.

All algorithms were designed and implemented as custom software written in Python version 3.14.6. Code was deployed and optimizations conducted on high-performance compute nodes utilizing Intel Xeon Platinum 8356H CPU and Nvidia A100 GPU processors.

## 3. Results

Here, we evaluated the performance of a proposed active learning protocol across three distinct scenarios, namely one where only sparse training data is available (N=10, representing 12.5% of the dataset), one where a moderate amount of data is available (N=30, 37.5%), and one where data is relatively abundant (N=60, 75%). In each case we performed extensive hyperparameter optimization (200 trials per model family) to identify the ideal combination of model architecture, training configuration and acquisition strategy. In addition to variability inherent in sample subset selection, one should must also consider the expected variability in class assignment induced by conformational variability. The latter represents an inherent replicate error that will introduce a universal upper bound on classification accuracy independent of any particular choice of model architecture, training strategy, or acquisition policy.

*3.1. Maximum Achievable Classification Accuracy*

Each epitope's primary sequence can lead to multiple 3D conformations. When conducting 10 repeated docking experiments with alternate conformations obtained for the same epitope RYKHTWGFE we obtain a range of affinity scores from -131.2 to -88.0 that provide a basis for estimating classification noise arising from replicate error (**Table 6**). Testing these 10 replicate docking affinity values, we find that they adhere approximated to a normal distribution with a mean value of -114.36 and a standard deviation of 11.63 (Shapiro-Wilk $p=0.29>0.05$ does *not* deviate from a normal distribution). Using this estimate and the Monte Carlo protocol described in Section 2.11, we find that this corresponds to an upper-bound in the classification accuracy of 78.7%. A close examination of the data points to conformation 6 as an outlier with a binding affinity value of -88, a substantially weaker binding score than the remaining conformations (>2 standard deviations away from the mean). As a result, we remove conformation 6 and we repeat the analysis. The estimated sample standard deviation decreases to 7.46 (Shapiro-Wilk $p=0.89 >>0.05$ or essentially a normal distribution), and the corresponding Monte Carlo estimate of maximum attain-able classification accuracy increases to 84.9%.

We further assess the results obtained for epitope RYKHTWGFE to determine if this is representative of the broader data set. Using this sequence in an additional set of 5 example sequences each with duplicate docking experiments performed we test for homoscedasticity. We find Bartlett's test $p\sim0.17$ and Levene's $p\sim0.15$ suggesting that replicate conformational error is not significantly non-uniform or heteroscedastic ($p>0.05$). If we focus specifically this second set of 5 example sequences assessed in duplicate we find that despite the much higher duplicate variance for sequence RYKHTWGFE, and the low degree of freedom available, replicate conformational error may again be considered reasonably uniform (homoscedastic) or at least not significantly heteroscedastic ($p>0.05$) (Bartlett's test $p\sim0.65$; Levene's $p\sim0.32$). Using the average standard deviation of these 5 duplicate pairs (Std Dev = 6.4), we repeat the same Monte Carlo corruption analysis described above. The resulting estimate of maximum attainable classification accuracy is 83.5%, aligning well with the outlier-corrected estimate obtained using 10 conformations for the primary sequence RYKHTWGFE

### *3.2. Architecture and Hyperparameter Stability*

Across all data availability regimes tested, the Transformer-based models consistently emerged as the highest-performing model architecture, outperforming optimal configurations of CNN, MLP, and the Linear baseline models (**Table 8)**. This result suggests that the self-attention mechanism may be uniquely superior at capturing the sequential token-like nature of amino acid in the 9-mer epitopes, even when labeled data is extremely scarce. Analysis of the optimal hyperparameter assignments reveals distinct trends in stability and adaptation. First, the use of ensembles was universally beneficial; every top-performing configuration utilized an ensemble of between 13 and 16 models. This is consistent with the expectation that ensembles reduce prediction variability. Interestingly, even though we expect data sparse regimes to be particularly vulnerable, an ensemble configuration is still selected even when 75% of the data is made available for training. Secondly, training to predict a smooth normalized affinity target (affinity proxy training) using MSE or Huber loss and then assigning the result to a class consistently outperformed training directly on the discrete class assignments. This result suggests that a more detailed supervision signal, effectively expands the information content of each training sample leading to more accurate subsequent class assignment. We also observe a counterintuitive inverse relationship between data availability and optimal model size with the optimal embedding dimension decreasing from 256 to 48 as the training set grows from N=10 to N=60. In the case of extremely sparse data, it would appear that the optimizer favors a more richly parameterized configuration where high-dimensional projections allow the separation of the few available training samples, as opposed to larger datasets which provide more constraints to train more compact and generalizable models. Conversely, while the Gaussian Error Linear Unit (GELU) is the standard activation function for modern Transformers, our optimization found that LeakyReLU was superior when only very sparse data is available for training(N=10). In this case, complexity may outweigh granularity. Indeed, one might expect the simpler, piecewise-linear gradient of LeakyReLU to support a more robust optimization, whereas the complex non-monotonic gradient profile of GELU may introduce unnecessary optimization difficulty when the supervision signal is sparse. Accordingly, as data availability increases to N=30 and N=60 training examples, the optimal configuration shifts to a GELU activation function where non-linearity of the latter better captures more complex binding relationships.

### 3.3. Active Learning Strategies and Training Trajectories

The strategy for optimal training shifts significantly as the budget for labeled data increases. For the minimal case of N=10 (12.5% of the dataset), the optimization converges to a one-shot training approach ($N_{\text{init}}$=10 of 10) with a predicted classification accuracy of roughly 72%, as there is insufficient data to support a meaningful split between an initial training set and a candidate acquisition pool. With a slightly larger initial training set of N=30, performance remains very close to a one-shot training, ($N_{\text{init}}$=26 out of 30) with the optimal choice of only one "active" learning step of $k_{\text{step}} = 4$ samples exhausting the full available data set. As a result, the choice of acquisition policy becomes irrelevant, with EGL being selected due to regularization or simply noise in the hyperparameter optimization process where for example $N_{\text{init}} = 26$ and $N_{\text{init}} = 30$ likely produce extremely similar performance, with predicted average classification accuracies of ~79% and ~81% respectively. However, as the data budget expands, the benefit of selecting an optimal acquisition policy for active incremental learning becomes clear. At N=60 (75% of the dataset), the best performing strategy involves a "slow drip" acquisition where the model is initially trained with a small set of 10/60 samples and where we incrementally add only $k_{\text{step}} = 1$ sample at a time using the Expected Gradient Length (EGL) selection policy (**Figure 6(A)**). While this strategy results in a high number of retraining steps and therefore higher computational runtime, it allows the model to better refine the decision boundary, using the single most informative sample at each learning stage. Indeed, the EGL policy consistently outperformed Random, Least Confident, and K-Center acquisition in these optimized runs. Selecting samples that induce the largest expected change in the model parameters appears to better mitigate the risk of overfitting to the initial random split, offering a more reliable proxy for "informativeness" than simple geometric coverage or output uncertainty.

### 3.4. Optimization Efficiency and Model Family Comparison

To assess the relative difficulty of finding these high-performance configurations, we analyzed the progression of the best-found accuracy as a function of the number of optimization trials for different model families up to a maximum allowable 200 trials (**Figure 7**). In the initial phases of optimization, CNN architectures are competitive in performance with the Transformer model. However, as the number of optimization trials increases, the Transformer family consistently converges to a higher classification accuracy, exceeding the best CNN configurations by margins of 2–5%. In contrast, the Linear and MLP models converge early at significantly lower accuracies, suggesting that simple positional weighting or fully connected layers are insufficient for modeling the complex non-linearities of epitope binding.

To further quantify the performance benefits of the proposed framework, we attempt to define a representative "naive" or "typical " baseline design. To avoid an inherently subjective choice of architecture (e.g., simple CNN) or default training parameters (e.g., learning rate of 1e-3, batch size 64) we construct a "Standard Benchmark" that we consider to be much more conservative. For this purpose, we select the optimal model configuration discovered using N=30 available training examples, namely a Transformer model with optimal learning rate, weight decay, and dropout settings (**Table 8**). We then make unavailable the advanced settings, namely ensemble solutions are disabled (single model training) and active learning is replaced with a one-shot learning of the full training allowance. This more conservative benchmark corresponds to a highly tuned best-case scenario for a standard machine learning approach, as it benefits from the model architecture produced by the TPE parameter search. The results of this comparison highlight a key efficiency threshold (**Table 9**) with the combined optimization of model architecture and learning strategies delivering a higher classification accuracy (80.5%) with only 30 samples than that delivered by the Standard Benchmark utilizing 60 samples (78.0%). This suggests that ensemble-based active learning and hyperparameter optimization contribute substantially to improvements in classification accuracy, surpassing the performance of a standard supervised approach while using half the amount of labeled data. It is important to emphasize that this comparison is conservative. The Standard Benchmark uses a Transformer architecture and training hyperparameters that were mathematically optimized to maximize performance at N=30, then applied to N=10 and N=60. A truly naive manual design lacking the specific balance of high weight decay and low dropout discovered by the automated search would almost certainly yield even lower accuracy. The widening gap in performance at N=60 (8.8%) training examples further suggests that standard training methods struggle to leverage additional data as effectively as the active incremental strategy, which specifically utilizes the most informative samples in a deliberate way.

### *3.5. Performance Relative to the Conformational Noise Ceiling*

The optimized classifier architecture and learning strategy achieves a peak classification accuracy of 86.8% when up to 75% of the data (N=60) is available. This result must be interpreted in the context of the noise ceiling estimated in Section 2.11. Our Monte Carlo simulations, which mimicked the effects of conformational variability on docking affinity, placed the theoretical maximum achievable accuracy between 83.5% and 84.9%. Convergence of the optimized model's classification accuracy at roughly the same values suggests that the combination of Transformer architectures, ensemble learning and EGL-based acquisition has effectively extracted nearly all accessible signal from the dataset. Indeed, maximum discriminatory performance is achieved with N~58 incrementally selected training examples, yielding optimal area under the curve (AUC) values of 0.891 for the prediction-recall curve (PRC) and 0.918 for the receiver-operating characteristic (ROC) curve (**Figure 6(B); Supplementary Figure 1**). This coincides with our estimates of a theoretical bound of ~85% beyond which the model is operating at the limit of what is biologically distinguishable and overfitting to conformational noise.

## 4. Discussion

In this work, we demonstrated that the computational screening of PRRSV vaccine epitopes into strong or weak candidates can be accelerated by adopting a model architecture and active learning strategy that is optimally aligned with this specific task. Moreover, we show that increases in model robustness resulting from such optimal design choices can effectively counteract the otherwise detrimental effects of data sparsity. For example, optimal hyperparameter tuning supported a higher classification accuracy using only 37.5% of the dataset (N=30) than did a conservative standard supervised approach using 75% of the dataset (N=60). When additional data is made available, an optimized transformer model reported an accuracy of 86.8%, slightly exceeding our estimates of the theoretical ceiling imposed by conformational variability. This peak performance was achieved at a practical saturation point of N~58 incrementally selected training examples, indicating that the model successfully captured the full self-consistent signal available within this dataset. Indeed, when only 10 or 30 samples were made available, essentially all samples were selected to populate the initial training set ($N_{init}$=10 of 10 and $N_{init}$ = 26 of 30) and perceived as novel or informative, effectively forgoing any opportunity for incremental learning.

Comprehensively optimizing over a variety of model architectures and acquisition strategies, our results suggest that Transformer-based models, supported by ensemble decision rules that also leverage Expected Gradient Length (EGL) for active-learning sample acquisition, consistently outperform traditional CNN and MLP baselines. The ability of transformer models to better capitalize on context in reading amino acid sequences has been reported by others, with Bidirectional Encoder Representations from Transformers (BERT) architectures outperforming CNN and Support Vector Machine (SVM) models in screening for protein-protein docking affinity with classification accuracies very similar to those reported here (~85%) [Malviya et al., 2024]. Moreover, this advantage has been shown to hold true and become even more substantial when data are sparse [Chen et al., 2024]. Indeed, the most recent transformer-based protein-protein language models (PPLM) are pre-trained on millions of example protein pairs and are capable of delivering predicted affinity accuracies approaching 90% [Liu et al., 2026]. In contrast, the transformer models presented here were not pretrained and initial model weights were sampled from a random uniform distribution. Interestingly the inherent capabilities of this architecture are such that even our protein-naïve transformer model was able to deliver classification accuracies approaching those of larger pretrained models despite using much smaller learning sets.

Findings such as these further emphasize the key role of incremental learning and optimal data acquisition polices and the optimization thereof demonstrated here [Szymborski and Emad, 2026]. In keeping with this, optimal learning configurations found in this work favor training on the continuous binding affinity scores as opposed to the discrete classes directly, recognizing that by doing so we preserve the granularity and natural progressions in the data thereby providing a richer signal for gradient estimation [de Vries and Thierens, 2025]. Representation in the feature space of each amino acid sequence was also found to play an important role. Indeed, the optimal dimensionality of the embedding space was found here to be inversely proportional to the number of training examples available, with sparse training sets being represented in much higher-dimensional spaces. Augmentation of the feature space [DeVries and Taylor, 2017] has in fact been shown to disproportionately affect information content (entropy) over increases in sample set size [Naumov, 2019], essentially providing added semantic context to each example and by doing so favor the discovery of more robust decision boundaries from fewer labeled samples.

Though many of the settings selected for model structure and learning in the optimal configuration may appear intuitive in retrospect, there was no guarantee of a collective positive synergy emerging from their combined and integrated use. One may also argue that their focused use on very sparse datasets runs countercurrent to the popular trend favouring increasingly large datasets. In situations such as the rapid emergence of a pandemic threat from a novel poorly known pathogen, the need for robust analytics in very sparse data becomes clear. In this work we show the potential for rapid computational screening of vaccine antigens given appropriate domain-specific choices of model design and learning strategies. Though work in this area is ongoing, methodologies such as these continue to show promise in significantly accelerating the timeline from pathogen identification to epitope selection.

## Author contributions

AEB developed the numerical methodology, AEB, GB and ZK co-designed the experiments. AEB and GB co-authored the manuscript with the guidance of ZK, HLW, ST and SR. AEB wrote all optimization codes, conducted the experiments, analyzed the data. ZK conducted all docking experiments and generated all binding affinity estimates in the full data set. BB and SR supported access and use of computing clusters. GB and ST conceived the project and received funding for the project. All authors reviewed and approved the final manuscript.

## Acknowledgments

This work was supported by the University of Saskatchewan's Centre for Quantum Topology and Its Applications (quanTA) and by the Vaccine and Infectious Disease Organization (VIDO). VIDO receives operational funding from the Canada Foundation for Innovation (CFI) through the Major Science Initiatives Fund and from the Government of Saskatchewan through Innovation Saskatchewan and the Ministry of Agriculture. The quanTA Centre's high-performance computational work has been advanced through a CFI John R. Evans Leaders Fund grant. We thank the University of Saskatchewan Advanced Research Computing (ARC) team for their efforts and responsiveness in creating an excellent local environment for this computational work. This article is published with the permission of the Director of VIDO.

## Competing interests

The authors declare no competing interests.

## Data availability

The datasets generated and analyzed in this work, as well as Python scripts with example use case files are available upon request

## Supplementary information

Supplementary Table 1.

Supplementary Table 2.

Supplementary Figure 1.

| Model Family | Structure Parameter | Search Space |
|---|---|---|
| All (incl. Lin) | $p_{\text{drop}}$ dropout probability | $p_{\text{drop}} \in \mathbb{R}, 0 \leq p_{\text{drop}} \leq 0.5$ |
| Linear | $\sigma$, activation function | $\sigma \in$ {ReLU } |
| MLP | $\sigma$, activation function | $\sigma \in$ {ReLU, LeakyReLU, ELU, GELU} |
| | $L_{\text{MLP}}$, layer count | $L_{\text{MLP}} \in \{1, 2, 3\}$ |
| | $d_{\text{MLP}}$, hidden layer(s) dimension | $d_{\text{MLP}} \in \{16, 32, 64, 128\}$ |
| CNN | $\sigma$, activation function | $\sigma \in$ {ReLU } |
| | $C_{\text{CNN}}$, base channel count | $C_{\text{CNN}} \in \{8, 16, 32, 64\}$ |
| | $L_{\text{CNN}}$, convolutional depth | $L_{\text{CNN}} \in \{1, 2, 3\}$ |
| | $k_{\text{CNN}}$, kernel size | $k_{\text{CNN}} \in \{3, 5\}$ |
| | $\rho_{\text{CNN}}$, pooling operator | $\rho_{\text{CNN}} \in \{\text{max}, \text{average}\}$ |
| | $d_{\text{CNN,fc}}$, fully connected hidden layer dimension | $d_{\text{CNN,fc}} \in \{8, 16, 32, 64\}$ |
| Transformer | $\sigma$, activation function | $\sigma \in$ {ReLU, LeakyReLU, ELU, GELU} |
| | $d_{\text{Tr,model}}$, embedding dimension | $d_{\text{Tr,model}} \in \{32, 48, 64, 128, 256\}$ |
| | $h_{\text{Tr}}$, number of attention heads | $h_{\text{Tr}} \in \{1, 2, 4, 8, 16\}$ |
| | $L_{\text{Tr}}$, number of encoder layers | $L_{\text{Tr}} \in \{1, 2, 3, 4\}$ |
| | $d_{\text{Tr,ff}}$, feed forward dimension | $d_{\text{Tr,ff}} \in \{16, 32, 64, 128, 256, 512\}$ |

**Table 1.** Structural hyperparameters associated search space by model family.

| Category | Training Parameter | Search Space |
|---|---|---|
| Supervision | $\mathcal{T}$, training target | $\mathcal{T} \in \{\text{class}, \text{affinity}\}$ |
| Objective | $\mathcal{L}$, loss function | $\mathcal{L} \in \begin{cases} \text{BCE}, & \mathcal{T} = \text{class} \\ \text{MSE}, \text{MAE}, \text{Huber}, & \mathcal{T} = \text{affinity} \end{cases}$ |
| Optimization | $\mathcal{O}$, optimizer | $\mathcal{O} \in \{\text{Adam}, \text{AdamW}, \text{SGD}, \text{RMSProp}\}$ |
| | $L_r$, learning rate | $L_r \in \mathbb{R}, 10^{-4} \leq L_r \leq 10^{-2}$ |
| | $M$, SGD, RMSProp Momentum | $0 \leq M \leq 0.95$ |
| | $B$, batch size | $B \in \mathbb{Z}, 1 \leq B \leq N_{\text{train}}$ |
| Regularization | $\lambda_{\text{wd}}$, weight decay | $\lambda_{wd} \in \mathbb{R}, 10^{-6} \leq \lambda_{wd} \leq 10^{-2}$ |
| Schedule | $E_{\text{init}}$, initial training epochs | $E_{\text{init}} \in \mathbb{Z}, 1 \leq E_{\text{init}} \leq 24$ |
| Schedule | $E_{\text{step}}$, active learning per-step training epochs | $E_{\text{step}} \in \mathbb{Z}, 1 \leq E_{\text{step}} \leq 24$ |
| Active Learning | $k_{\text{step}}$, number of pool samples added to $D_{\text{train}}$ each active learning step | $k_{\text{step}} \in \mathbb{Z}, 1 \leq k_{\text{step}} \leq 8$ |

**Table 2.** Training hyperparameters and associated search space.

| Acquisition Policy | Description | Information Used |
| --- | --- | --- |
| Random | Uniform random selection from $\mathcal{D}_{\text{pool}}$ | None |
| Least Confident | Selects samples with predicted probability closest to 0.5 | Model output probabilities $p(x)$ on $\mathcal{D}_{\text{pool}}$ |
| K-Center | Greedy k-center selection in learned embedding space | Granular feature vectors $h(x)$ for $x \in \mathcal{D}_{\text{pool}}$ |
| EGL | Expected gradient length | $p(x)$ and penultimate feature vectors $h(x)$ for $x \in \mathcal{D}_{\text{pool}}$ |

**Table 2.** Acquisition policy descriptions and information used.

| Decision Rule | Description |
| --- | --- |
| Majority Vote | Majority vote over threshold probability |
| Mean Round | Thresholded mean predicted probability |
| Confidence Weighted | Vote weighted by distance from decision boundary |

**Table 3.** Ensemble decision rule choice descriptions.

| Category | Parameter | Search Space |
| --- | --- | --- |
| Active Learning | $N_{\text{init}}$, initial labeled set size | $N_{\text{init}} \in 2\mathbb{Z}, 2 \leq N_{\text{init}} \leq N_{\text{train}}$ |
| Active Learning | $\alpha$, acquisition policy | $\alpha \in$ {Random, Least Confident, K-Center, EGL} |
| Active Learning | $k_{\text{step}}$, acquisition batch size | $k_{\text{step}} \in \mathbb{Z}, 1 \leq k_{\text{step}} \leq 8$ |
| Ensemble | $e$, ensemble usage flag | $e \in \{0, 1\}$ |
| Ensemble | $M$, ensemble size | $M \in \mathbb{Z},\ 2 \leq M \leq 16$ if $e = 1$ |
| Ensemble | $\delta$, ensemble decision rule | $\delta \in$ {Majority Vote, Mean Round, Confidence Weighted} |

**Table 4.** Active learning and Ensemble hyperparameters.

| Epitope Sequence | Conformation | Docking score |
|---|---|---|
| RYKHTWGFE | 1 | -109.6 |
| RYKHTWGFE | 2 | -111.2 |
| RYKHTWGFE | 3 | -116.9 |
| RYKHTWGFE | 4 | -110.2 |
| RYKHTWGFE | 5 | -112.3 |
| RYKHTWGFE | 6 | -88 |
| RYKHTWGFE | 7 | -131.2 |
| RYKHTWGFE | 8 | -118 |
| RYKHTWGFE | 9 | -120.6 |
| RYKHTWGFE | 10 | -125.6 |

**Table 5.** Docking scores across conformations for a fixed epitope-receptor pair.

| Epitope Sequence | Method A Affin-ity | Method B Affin-ity | Variance $s^2$ |
|---|---|---|---|
| GVVNLVKYA | -62.6 | -56.5 | 18.6 |
| RYKHTWGFE | -91.6 | -108.2 | *137.8* |
| VKQGVVNLV | -57.9 | -60.3 | 2.9 |
| EFSLPTHHT | -93.7 | -101.2 | 28.1 |
| LIDLKRVVL | -61 | -67 | 18.0 |

**Table 6.** Affinity differences induced by alternative conformation selection methods.

| Training Sample Count / Parameter | N=10 (12.5%) | N= 30 (37.5%) | N=60 (75%) |
|---|---|---|---|
| Mean final test accuracy after hyperparameter optimization | 72.1% | 80.5% | 86.8% |
| Mean misclassified sample rate (100-accuracy) | 27.7% | 19.5% | 13.2% |
| $f_\theta$, Model Family | Transformer | Transformer | Transformer |
| $d_{\mathrm{Tr,model}}$, embedding dimension | 256 | 64 | 48 |
| $h_{\mathrm{Tr}}$, number of attention heads | 2 | 1 | 1 |
| $L_{\mathrm{Tr}}$, number of encoder layers | 1 | 3 | 2 |
| $d_{\mathrm{Tr,ff}}$, feed forward dimension | 512 | 32 | 64 |
| $p_{\mathrm{drop}}$, dropout probability | 0.151 | 0.022 | 0.371 |
| $\sigma$, hidden layer activation function | leaky_relu | gelu | gelu |
| $N_{\mathrm{init}}$, initial labeled set size | 10 | 26 | 10 |
| $\alpha$, acquisition policy | N/A | *EGL* | EGL |
| $k_{\mathrm{step}}$, acquisition batch size | N/A | 4 | 1 |
| $e$, ensemble usage flag | True | True | True |
| $M$, ensemble size | 14 | 16 | 13 |
| $\delta$, ensemble decision rule | Mean Round | Mean Round | Confidence Weighted |
| $\mathcal{T}$, training target | Affinity | Affinity | Affinity |
| $\mathcal{L}$, loss function | MSE | MSE | Huber |
| $\mathcal{O}$, optimizer | AdamW | Adam | AdamW |
| $B$, batch size | 2 | 21 | 42 |
| $\lambda_{\mathrm{wd}}$, weight decay | 3.09e-4 | 1.21e-6 | 5.09e-5 |
| $E_{\mathrm{init}}$ , initial training epochs | 18 | 28 | 17 |
| $E_{\mathrm{step}}$ , step training epochs | 4 | 29 | 17 |
| $\mathcal{T}$, training target | Affinity | Affinity | Affinity |

**Table 7.** Optimized hyperparameters and performance metrics by training sample count.

| Training Sample Count | Standard Benchmark Accuracy | Optimized Pipeline Accuracy | Performance Delta |
|---|---|---|---|
| N=10(12.5%) | 68.1% | 72.1% | +4.0% |
| N=30(37.5%) | 76.7% | 80.5% | +3.8% |
| N=60(75.0%) | 78.0% | 86.8% | +8.8% |

**Table 8.** Comparison of the full Optimized Pipeline vs. the Standard Benchmark

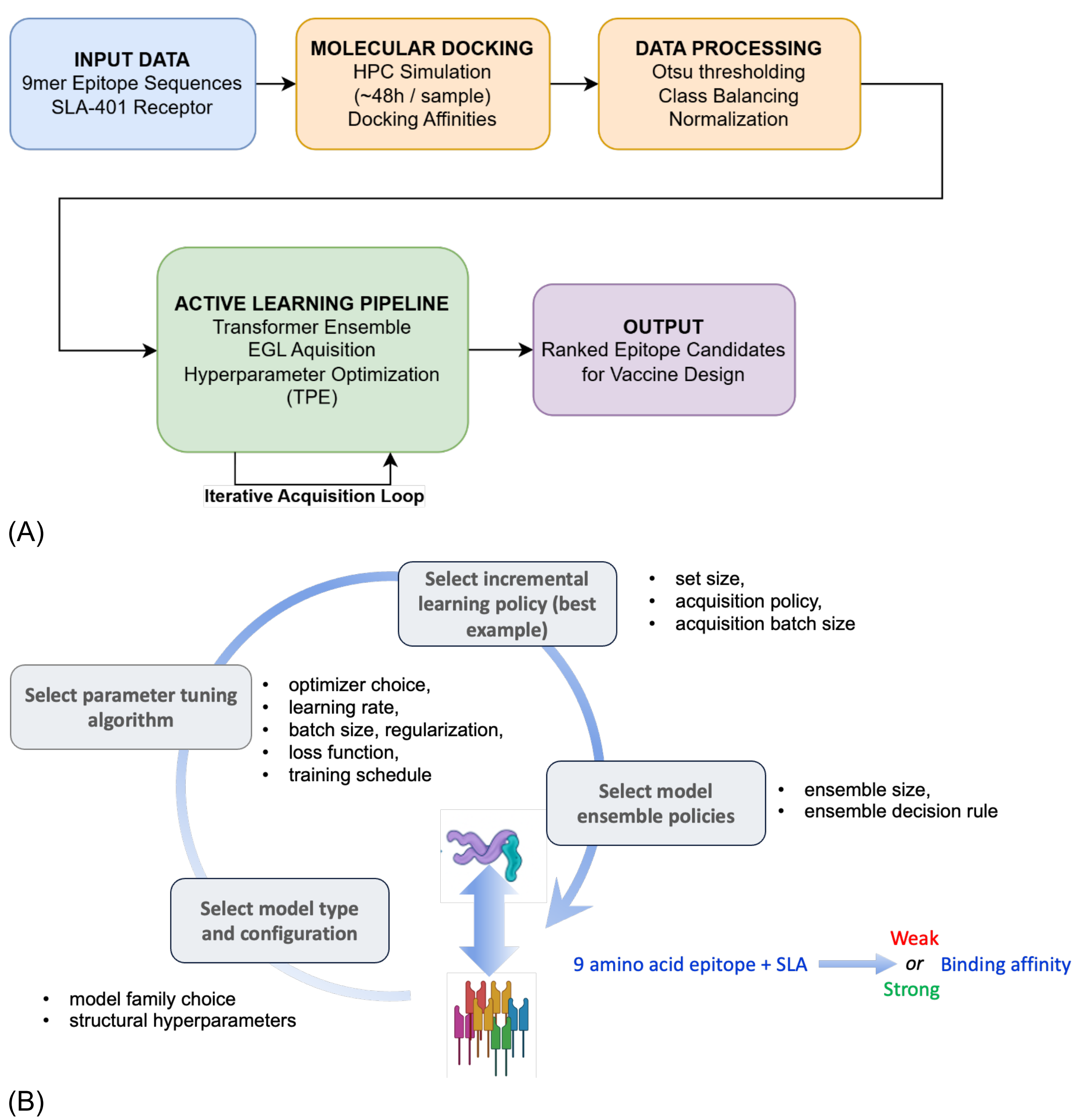

**Figure 1**. *Framework conceptual overview*. A graphical outline of the proposed analytical framework's data flow (A), with additional detail of the active learning and model architecture optimization component (B)

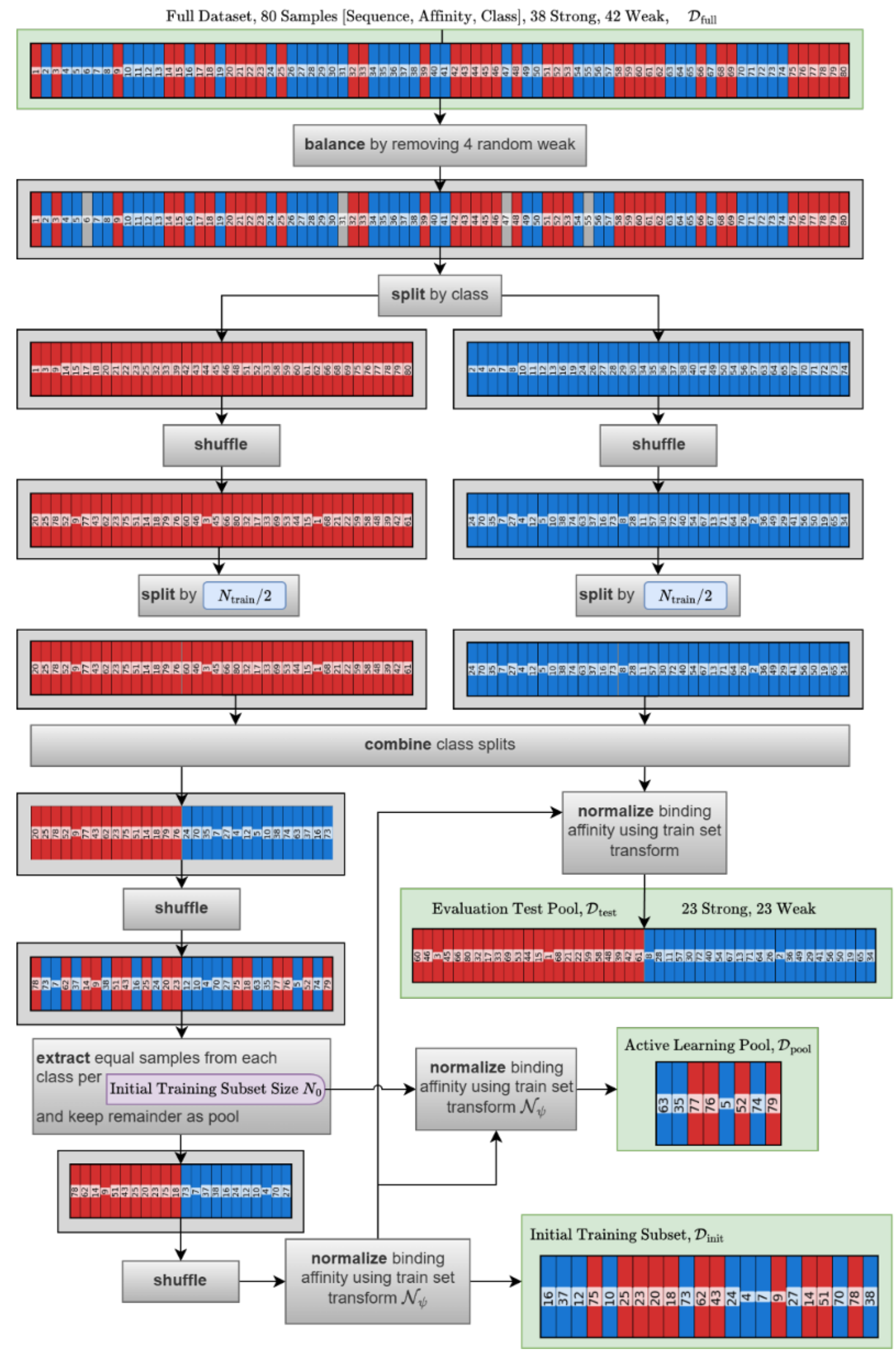

**Figure 2**. *Data preparation process*. Stepwise operations for data preparation applied to a sample size of N=30, and a pool size of 8.

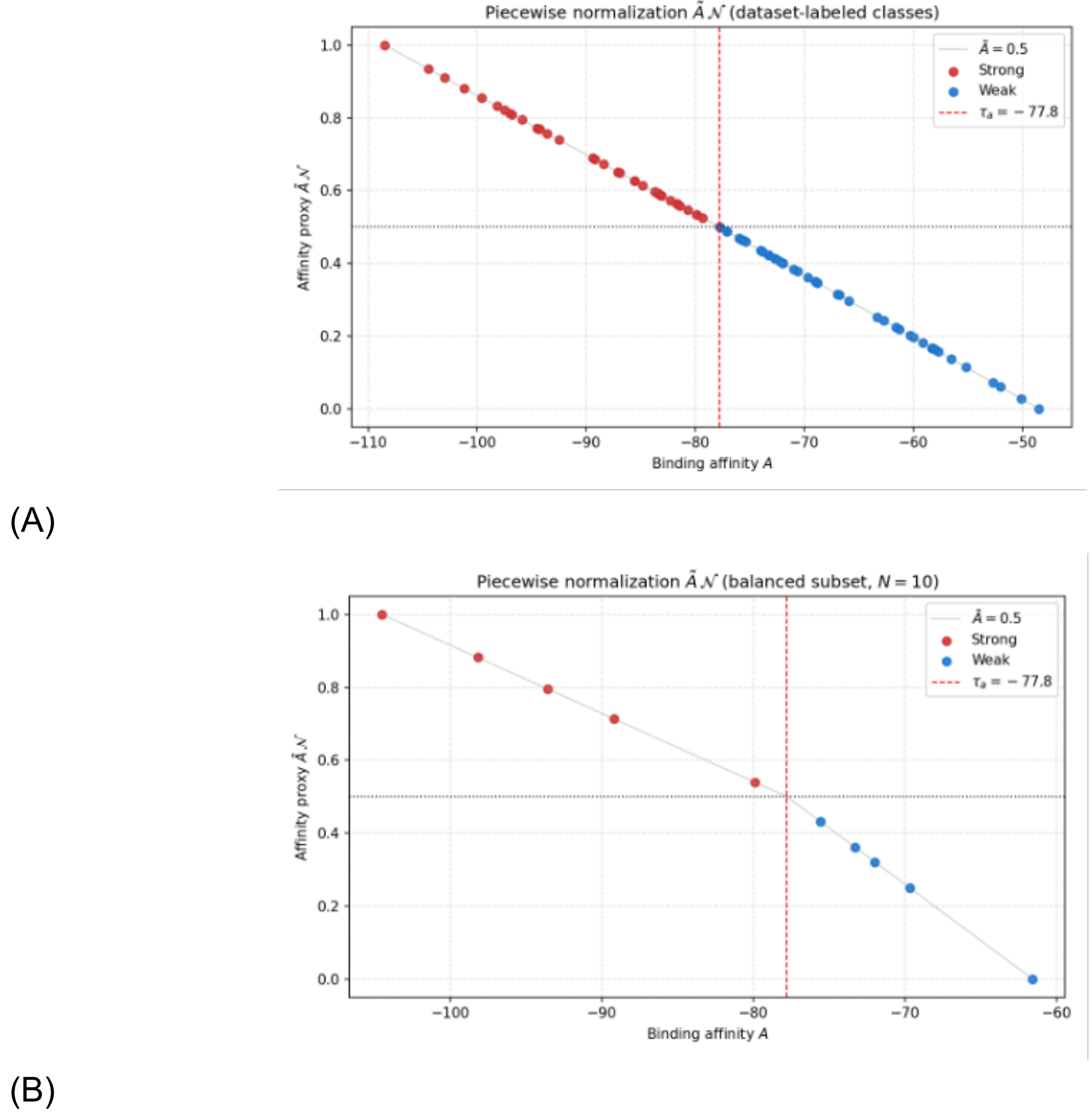

**Figure 3**. *Piecewise normalization results*. Piecewise normalization applied to the full dataset (A) and to a small random subset of the data, highlighting the dynamic range capability of this approach (B)

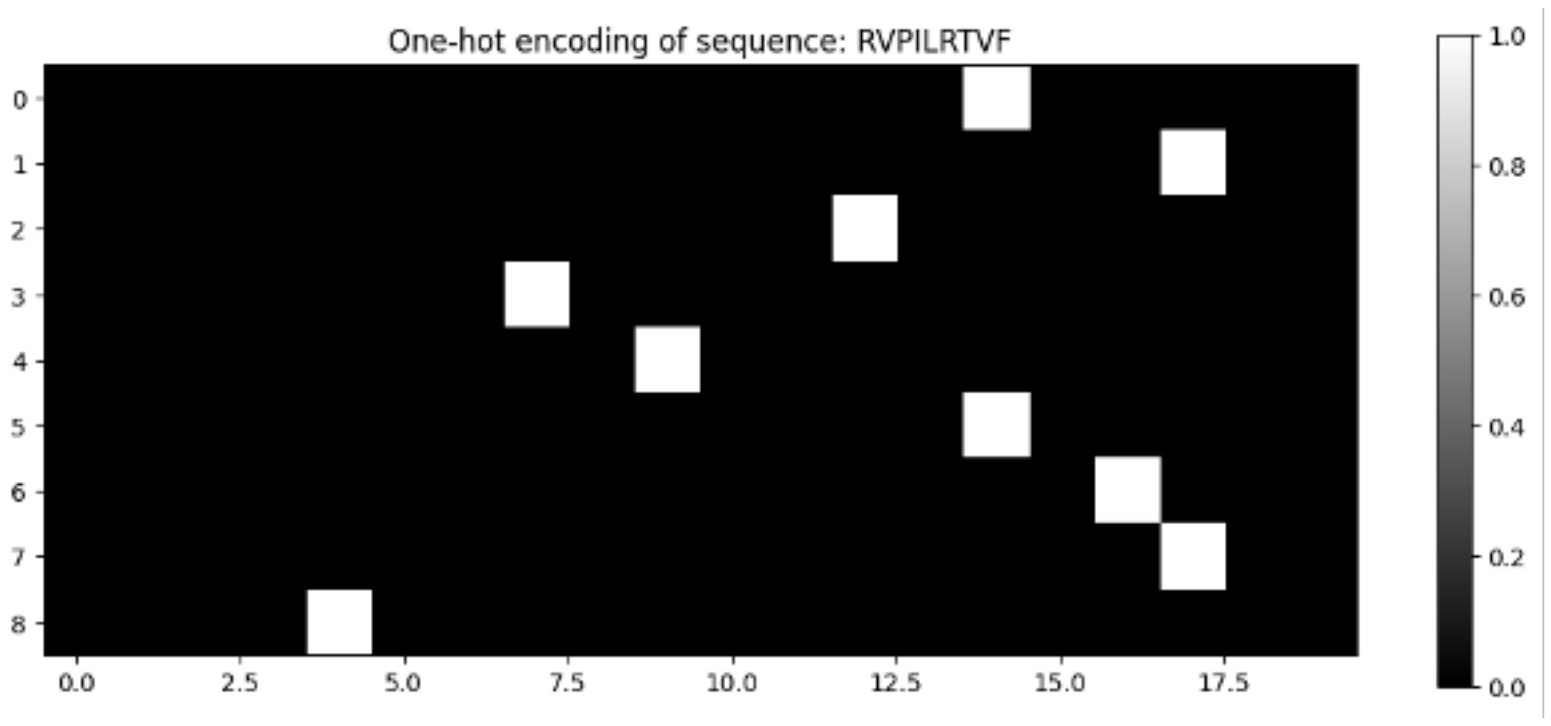

**Figure 4**. *One-hot encoding*. Example of a resulting one-hot encoding of the amino acid sequence RVPILRTVF

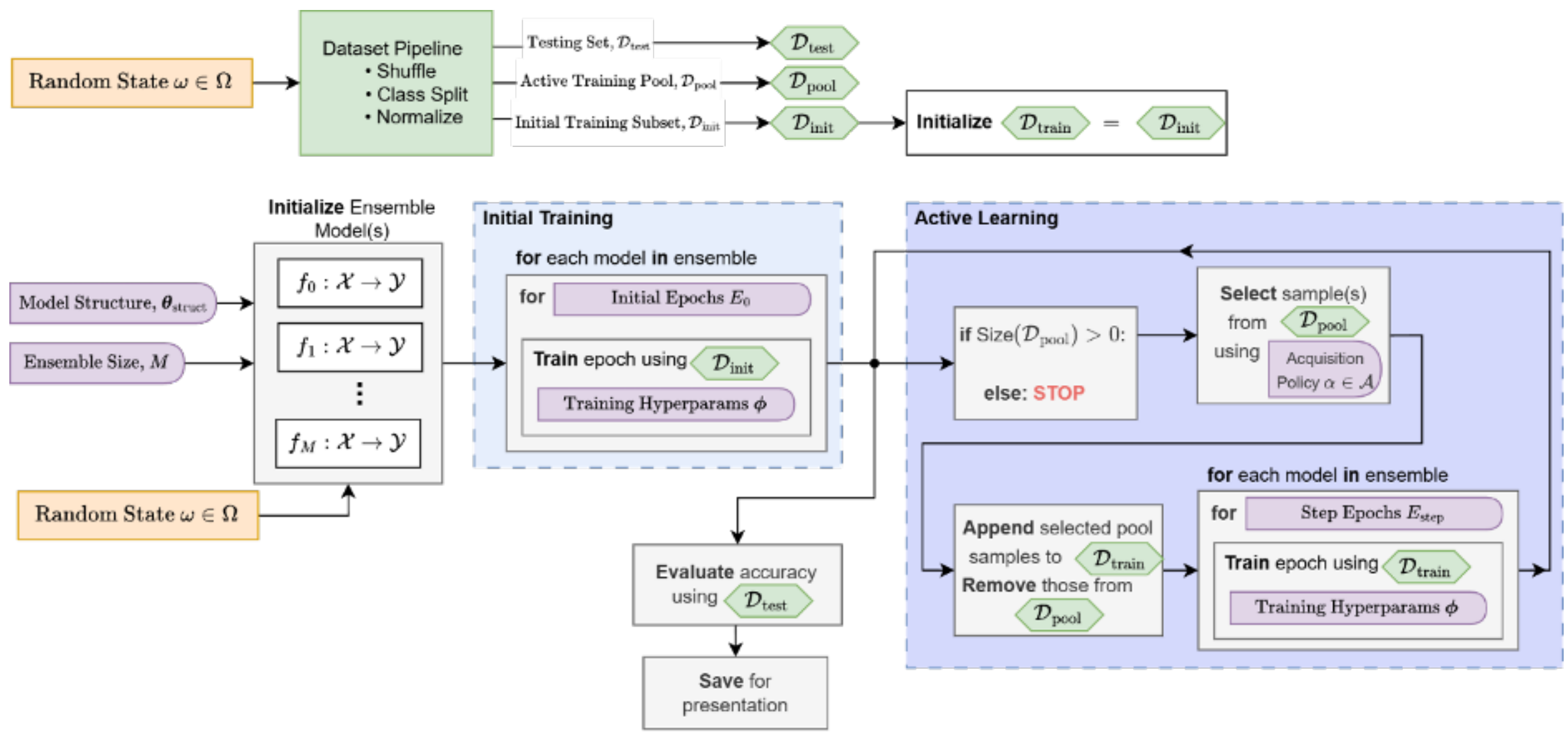

**Figure 5**. *Overview of active learning process*. Descriptive analytical flow of operations involved in a single iteration of the training strategy using active learning.

(A)

(B)

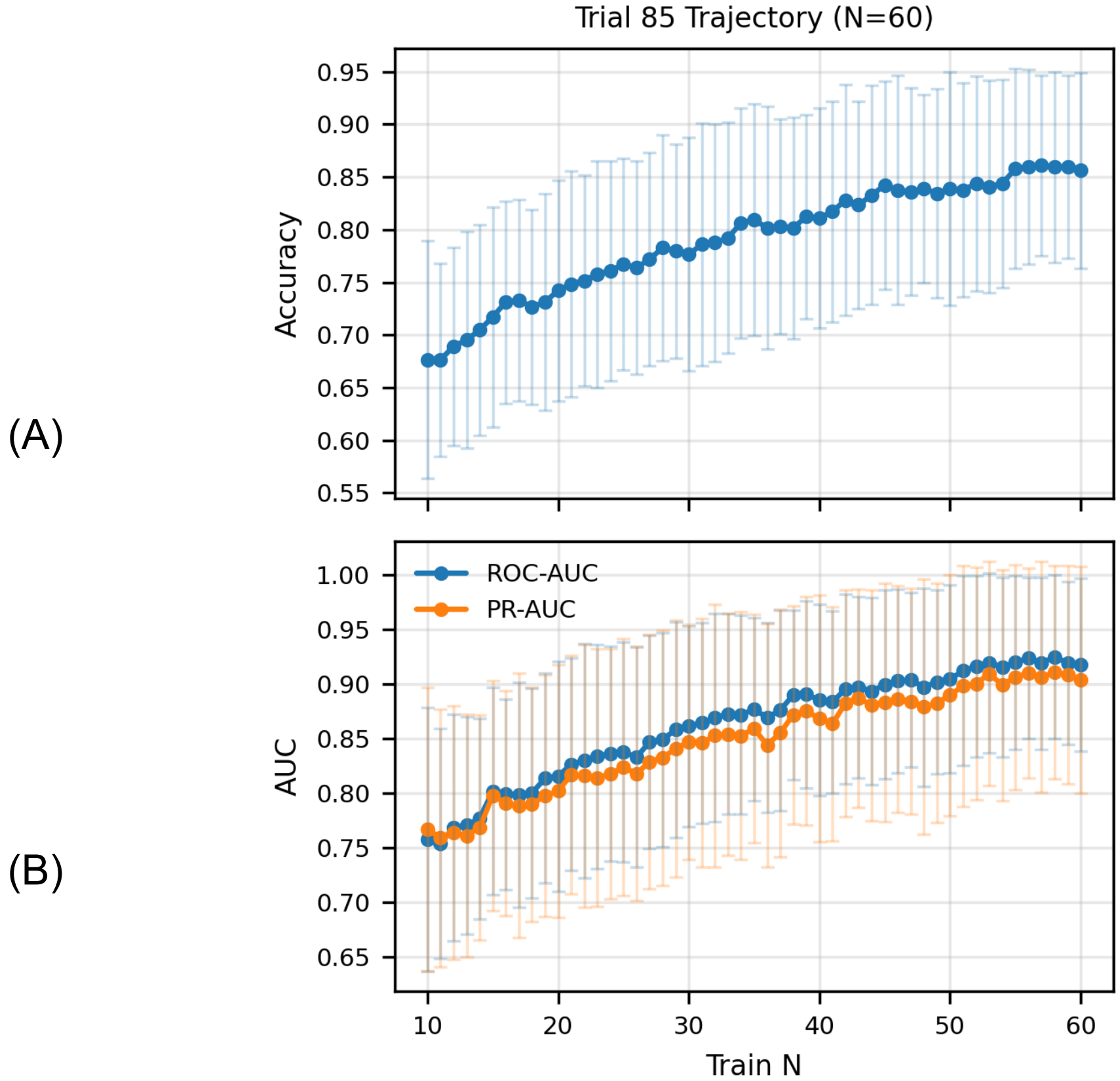

**Figure 6**. Optimal *active learning trajectory.* Trajectory of the prediction accuracy in the test set for an optimal configuration of active learning for an optimally configured Transformer model architecture using N= 60 out of 80 samples available for training (A). Corresponding changes to the area under the curve (AUC) for the Receiver Operating Characteristic (ROC) curve and Precision-Recall Curve (PRC) are shown in the lower panel (B). Data points represent the average and standard deviation of values across 40 repeat sample acquisition runs. Peak $AUC_{ROC}$ = 0.918 and $AUC_{PRC}$=0.891 are achieved at N=58.

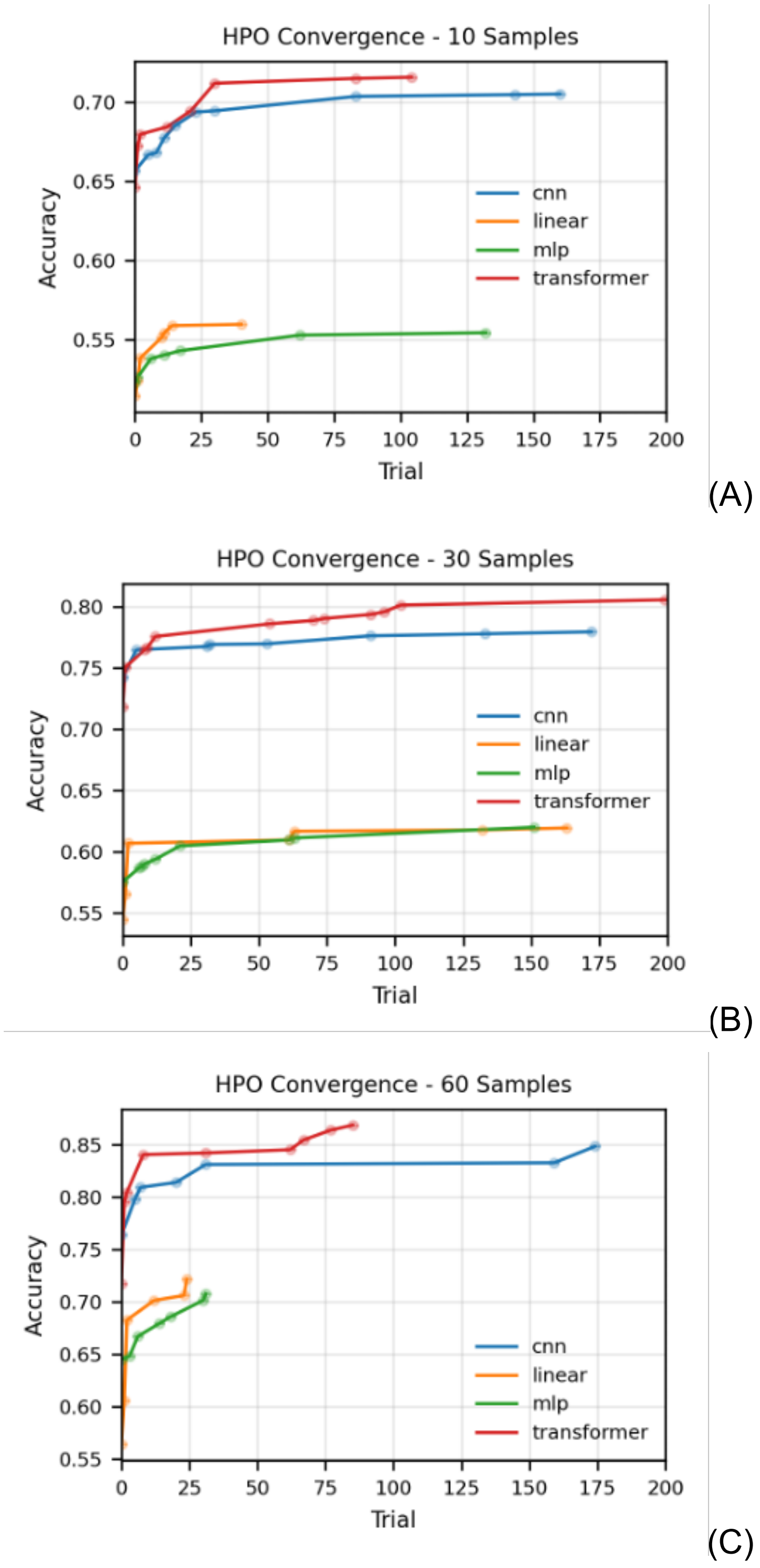

**Figure 7**. *Comparing model architectures*. Predictive accuracy in test set for different model architectures during incremental hyperparameter optimization using N=10(A), 30 and 60 training samples out of 80 available samples overall.

```
Inputs:
   data_file              // CSV file containing sequences and energies
   total_train_size    // fixed training set size
   initial_label_size  // number of initially labeled samples
   random_seed
Outputs:
   X_encoded           // one-hot encoded sequences
   labels                  // binary class labels
   init_indices          // initially labeled training samples
   pool_indices         // unlabeled pool samples
   test_indices          // held-out test samples
Procedure:
1. Load data from file
   data ← read_csv(data_file)

2. Encode sequences and read class labels
   for each row in data:
      X_encoded[i] ← one_hot_encode(sequence[i])
      labels[i] ← class_label_from_file[i]

3. Separate samples by class
   class0_indices ← indices where labels == 0
   class1_indices ← indices where labels == 1

4. Shuffle each class independently
   set_random_seed(random_seed)
   shuffle(class0_indices)
   shuffle(class1_indices)

5. Enforce strict class balance
   samples_per_class ← min(size(class0_indices), size(class1_indices))
   class0_indices ← first samples_per_class elements
   class1_indices ← first samples_per_class elements

6. Split each class into training and test subsets
   train_class0 ← first (total_train_size / 2) elements of class0_indices
   test_class0  ← remaining elements
   train_class1 ← first (total_train_size / 2) elements of class1_indices
   test_class1  ← remaining elements

7. Form balanced training and test sets
   train_indices ← concatenate(train_class0, train_class1)
   test_indices  ← concatenate(test_class0, test_class1)
   shuffle(train_indices)
   shuffle(test_indices)

8. Select initial labeled training set
   init_indices ← balanced subset of train_indices with size initial_label_size
   pool_indices ← remaining train_indices
   shuffle(pool_indices)

9. Compute energy normalization
   reference_energies ← energy values at init_indices
   normalization_function ← fit_normalization(reference_energies, energy_threshold)

10. Apply normalization for energy-based training
   for each index in train_indices:
      normalized_energy[index] ← normalization_function(energy[index])

Return X_encoded, labels, init_indices, pool_indices, test_indices
```

**Supplementary Table 2.** Data preparation procedure pseudo-code.

```
Inputs:
   X_encoded        // one-hot encoded sequences
   labels           // binary class labels
   affinities       // docking affinity values
   init_indices     // initially labeled training samples
   pool_indices     // unlabeled pool samples
   test_indices     // held-out test samples
   model_config     // model architecture configuration
   training_config  // optimizer, loss, epochs, batch size, AL parameters
   random_seed
Outputs:  test_accuracies  // test accuracy after initial training and each AL step
Procedure:
1. Set random state
   set_random_seed(random_seed)

2. Initialize training and pool sets
   train_indices ← init_indices
   pool_indices  ← pool_indices

3. Build test data loader
   test_loader ← build_loader(X_encoded, labels, affinities, test_indices)

4. Initialize model(s)
   if training_config.use_ensemble:
      for i = 1 to ensemble_size:
         models[i] ← build_model(model_config)
   else:
      models ← { build_model(model_config) }

5. Initial training
   train_loader ← build_loader(X_encoded, labels, affinities, train_indices, train_target)
   for each model in models:
      train(model, train_loader, init_epochs, optimizer, loss)

6. Initial evaluation
   test_accuracies[0] ← evaluate_accuracy(models, test_loader, ensemble_decision)

7. Active learning loop
   while pool_indices is not empty:
      k ← min(step_k, size(pool_indices))
      selected_indices ← select_samples(
                          selection_method,
                          models,
                          X_encoded,
                          train_indices,
                          pool_indices,
                          k)
      pool_indices  ← pool_indices \ selected_indices
      train_indices ← train_indices ∪ selected_indices
      train_loader ← build_loader(X_encoded, labels, affinities, train_indices, train_target)
      for each model in models:
         train(model, train_loader, step_epochs, optimizer, loss)

      append evaluate_accuracy(models, test_loader, ensemble_decision) to test_accuracies

8. Return results
   return test_accuracies
```

**Supplementary Table 3**. Active-learning training strategy pseudo-code

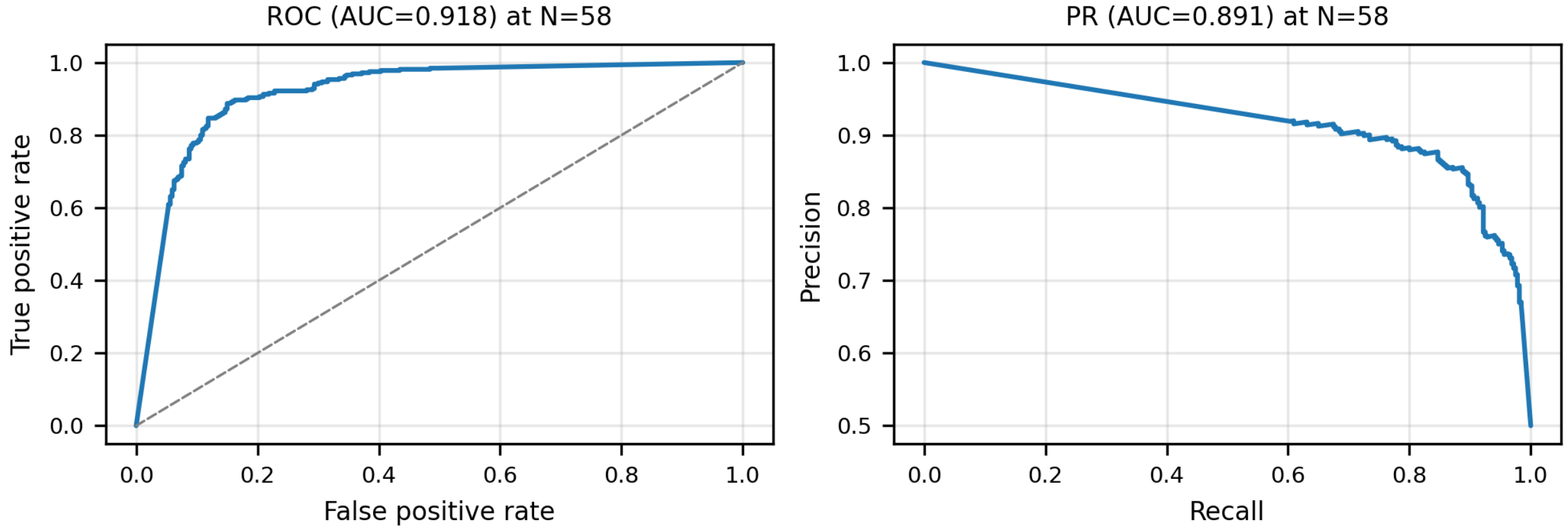

**Supplementary Figure 1**. *Classification performance*. Maximal area under the curve (AUC) values of 0.891 for the prediction-recall curve (PRC) and 0.918 for the receiver-operating characteristic (ROC) curve achieved by an optimally configured transformer model with incremental recruitment of N=58 select training examples.